# Pulsed Neutron Spectroscopy of Low Dimensional Magnets: Past, Present, and Future

S. E. Nagler[1] and D. A. Tennant[2,3]

[1] Neutron Scattering Division, Oak Ridge National Laboratory, Oak Ridge, TN, USA
[2] Materials Science and Technology Division, Oak Ridge National Laboratory, Oak Ridge, TN, USA
[3] Shull-Wollan Center, Oak Ridge National Laboratory, Oak Ridge, TN, USA

E-mail: naglerse@ornl.gov; tennantda@ornl.gov

**Abstract**

The early 1990's saw the first useful application of pulsed neutron spectroscopy to the study of excitations in low dimensional magnetic systems, with Roger Cowley as a key participant in important early experiments. Since that time the technique has blossomed as a powerful tool utilizing vastly improved neutron instrumentation coupled with more powerful pulsed sources. Here we review representative experiments illustrating some of the fascinating physics that has been revealed in quasi-one and two dimensional systems.

Keywords: low dimensional magnetism, neutron scattering, fractional excitations

**Preface**

Professor Roger A. Cowley was an outstanding influence on, and a formal or informal mentor to both of the co-authors. Roger was the initial catalyst for our own collaborations which have now spanned nearly three decades, starting when one of us (SEN) joined his group for a sabbatical year at Oxford and the other (DAT) was beginning his Ph.D. research. Roger's great physical insights were inspirational, and his collaborative spirit set a great example for those of us working research problems together. We are pleased to contribute this article to the special journal issue honouring his legacy. Here we review selected time-of-flight inelastic neutron scattering studies of excitations in low dimensional magnets, principally those carried out by alumni of Roger's group or his close associates. Looking back on this body of research illustrates both how our thinking on the subject has evolved, as well as the great advances in neutron spectroscopy using pulsed spallation source that have made this possible.

## 1. Introduction

At one time low dimensional magnetic systems (i.e. of dimension $D = 1$ or $2$) were considered to be "toy" problems useful to the extent that solving them provided insights into the "real world" of $D = 3$. The realization [1] that the effects of both quantum and thermal fluctuations and non-linear behaviour were greatly enhanced in low dimensions motivated the study of these systems in their own right. A pair of reviews published in the 1970's gives a nice survey of work done up to that time on model magnetic systems in various dimensions [2] and in particular 1D [3].

Inelastic neutron scattering (i.e. neutron spectroscopy) has always played a unique role in this research since it provides a straightforward measurement of the magnetic excitations in the form of the space and time dependent dynamical susceptibility. For many systems this enables the determination of the effective magnetic Hamiltonian. Early measurements on the $S = 5/2$ Heisenberg antiferromagnetic chain (HAFC) material TMMC [4] showed well-defined spin waves at low temperatures with strong dispersion of the modes only in the direction of the chains (see figure 1). This was a clear illustration of one of the salient characteristics of low dimensional systems: correlations do not depend on the irrelevant dimensions. Thus scattering in simple quasi-1D systems is sheet-like, and scattering in quasi-2D systems is rod-like. As noted below neutron scattering experiments can be arranged to take advantage of these features.

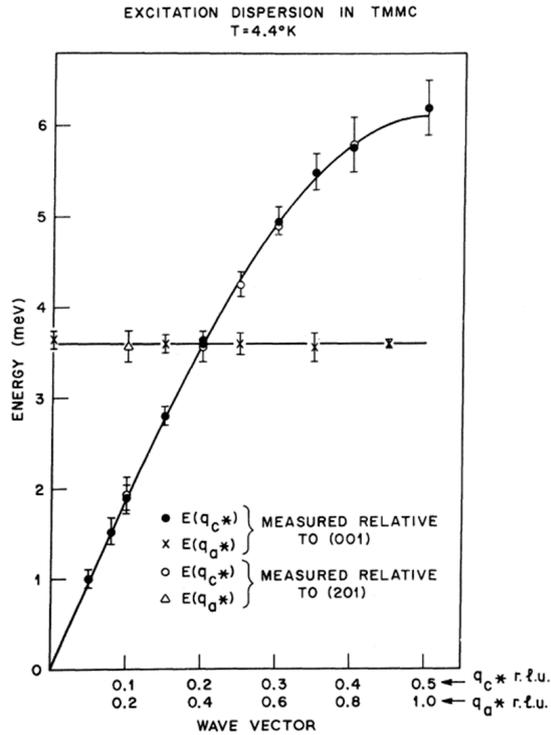

**Figure 1.** Spin wave dispersion in TMMC contrasting the strong energy dependence on momentum transfer along the chain direction (circles) with the independence on momentum perpendicular to the chain (from [4]).

Experimentally systems with spin quantum number S = 5/2 are semi-classical and neutron scattering measurements are well-described by modified classical theories. However, this is not true in the quantum limit, which turns out to be much more interesting. Historically, the ground state of the HAFC with Hamiltonian $2J\sum \vec{S}_r \cdot \vec{S}_{r+1}$ was solved exactly for S=1/2 using Bethe Ansatz techniques as early as 1931 [5]. Identifying the excited states was much more difficult, and little progress was made for 30 years until des Cloizeaux and Pearson (dCP) identified a set of low lying spin-wave like states [6]. The prevailing wisdom in the early 1970s was that the principal quantum effect was simply a renormalization of the spectrum: the classical expression for the spin wave energy, $E_q^{SW} = 2J|\sin(q)|$ was replaced for S=1/2 by the relation $E_q^{dCP} = \pi J|\sin(q)|$. Neutron measurements of the S = 1/2 HAFC material CPC initially confirmed the applicability of the dCP spectrum [7], but detailed examinations of the scattering [8] showed asymmetric lineshapes with significant spectral weight at higher energies, particularly in the vicinity of $q = \pi$ (see figure 2). This additional intensity was expected to some extent based on analytical [9] and finite chain [10] calculations of the spectrum but its true physical significance was not yet apparent.

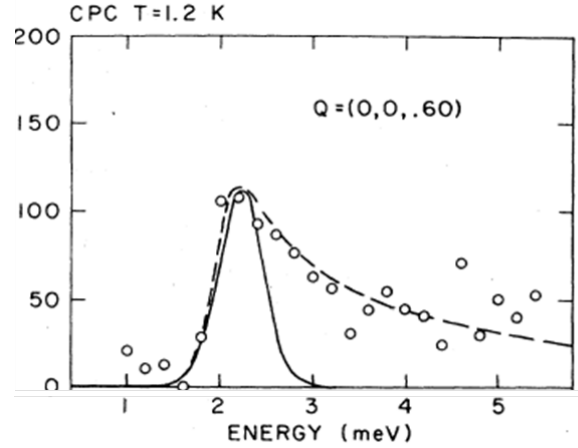

**Figure 2**. Lineshape measured in CPC for $q = \frac{6}{5}\pi$. The solid line shows the expected instrumental resolution limited lineshape. (Adapted from [8]).

Our general understanding of the physics of excitations in the HAFC was revolutionized starting in the 1980's. As is now widely known, the fundamental excitations are fractionalized spinons, with a spin S = 1/2 relative to the singlet ground state [11]. As shown by Haldane [12], when the underlying spin S is an integer (e.g. S = 1) the spinons are bound and the resulting excitation spectrum is dominated by S = 1 triplons. The periodicity of the excitations extends over a momentum range of $2\pi$ as expected for a system without magnetic order, and exhibits a gap (the "Haldane gap") at the antiferromagnetic vector q = π. The initial experimental observation of these features for S = 1 chains came in 1986 via measurements on the material $CsNiCl_3$ [13]. For S = 1/2 HAFCs there was substantial theoretical activity aimed at calculating the dynamical suscpetibility. It was determined that the free spinon spectrum had a lower bound at the dCP energy, and a momentum dependent upper boundary $E_q^{UB} = 2\pi J\left|\sin\left(\frac{q}{2}\right)\right|$. The best prediction for the dynamical suscpetibility was the "Müller ansatz" based on existing calculations as well as selection rules [14]. An approximate analytic expression for the triply degenerate dynamical correlation function $S^{\alpha\alpha}(Q,\omega)$ was proposed as a continuum with a square-root singularity at the lower bound given by $E_q^{dCP}$ and a sharp cutoff at the upper bound $E_q^{UB}$. The experimental situation remained much less clear. As discussed in the next section, pulsed neutron scattering would now play a crucial role.

## 2. Studies of spinons using MARI

Through the 1980's it was widely believed that for energy transfers less than 100 meV time-of-flight (ToF) chopper spectrometers were advantageous only for inelastic neutron scattering measurements where the scattering cross-section is isotropic in momentum space. Therefore chopper spectrometers were mainly used for studies of liquids, polycrystalline materials, incoherent inelastic scattering or localized excitations such as crystal field levels. Inelastic scattering in anisotropic systems was the province of triple axis spectrometry. The availability of improved computational power and new intense pulsed neutron sources, notably ISIS, led to a reconsideration of the situation. Toby Perring in particular began exploring how to best utilize direct geometry chopper spectrometers for investigations of single crystal dispersive excitations in 3D [15], and developed the eponymous Tobyplot as an aid in experimental planning. At the same time, it became clear that at least in principle there could be a large gain for ToF measurements of low D systems since with substantial detector coverage the signal could be summed over the irrelevant directions in reciprocal space without sacrificing important information. This concept was put to use in studies of the classical (1.e. S=5/2) 1D antiferromagnet $KFeS_2$ [16,17].

Mapping out a complete spectral function for the S=1/2 HAFC with a chopper spectrometer had occurred to one of us (SEN) following discussions with Rob Robinson when he was the instrument responsible for the constant-Q spectrometer at Los Alamos. The material $KCuF_3$ was available in large single crystal form and its magnetic excitations had been investigated using triple axis methods [18, 19]. The dispersion along the chain fit well to a sinusoidal dispersion with a zone boundary energy of 55 meV. Satija *et al.* [19] pointed out that in fact, both the susceptibility and dispersive excitations could be decribed in a consistent fashion by either classical or quantum (i.e. dCP) theory using a single exchange constant, and the same was true for CPC. Without apriori knowledge of the exchange constant it would be necessary to measure the full dynamic susceptibility and the form of the continuum scattering to definitively establish the presence of quantum effects. One expected the continuum in $KCuF_3$ to extend to energies above 100 meV, a range that is well suited to chopper spectrometers. A golden opportunity to pursue this idea arose in 1990 when SEN arranged a sabbatical year to be spent visiting Roger Cowley's group at the Oxford Clarendon Laboratory. A proposal to use the direct geometry MARI spectrometer [16] at the nearby ISIS pulsed neutron source was accepted. Shortly afterwards DAT joined the experiment as part of his graduate thesis work.

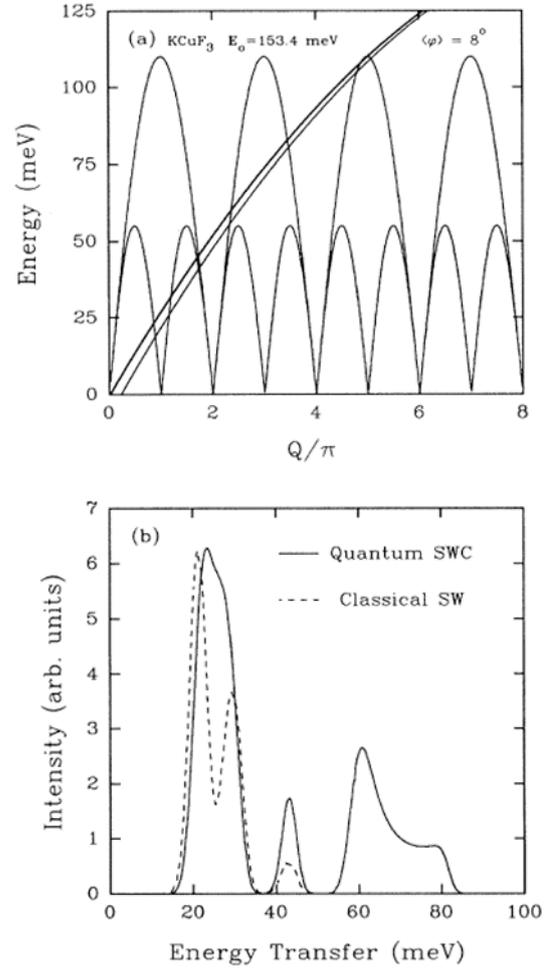

**Figure 3.** (a) Measurement trajectory in the MARI low angle bank in 1D momentum-energy space, with an incident energy $E_0 = 153.4$ meV and $k_i$ along c*. The dCP and upper bound energies are plotted in the extended zone scheme. The sampled phase space is the region between the two parabolic curves. (b) Simulated scattering for classical spin-wave theory (dashed line) and the quantum Müller ansatz (solid line). Figure from Ref. [20].

The early measurements at MARI were quite challenging as one had to deal with large backgrounds from the sample containers as well as significant multi-phonon scattering from the crystal itself. To maximize the magnetic signal the initial experiments [20] were carried out with the incident neutron wavevector, $k_i$, parallel to the magnetic chain direction (the tetragonal c* axis). In this configuration the signal in the low angle detector bank could be summed up while preserving the 1D magnetic scattering. Each detector sweeps out a parabolic trajectory in the 1D momentum-energy space (see figure 3(a)). The non-magnetic sample background was estimated by measuring with $k_i$ perpendicular to c* and selecting the detectors where the component of momentum transfer **q** along the chains was zero. Programs were written to analyze the data including instrumental resolution, and validated by comparison to the spin-wave scattering of $KFeS_2$. This ultimately enabled a direct comparison of the measured scattering with the Müller ansatz for the S=1/2 HAFC cross-section. Figure 3(b) shows the simulated scattering. The Müller ansatz was found to give an excellent description of the scattering throughout the Brillouin zone (figure 4). Purely quantum effects were manifestly evident by the presence of scattering in the energy region well above the zone boundary, with a cutoff at the expected upper bound.

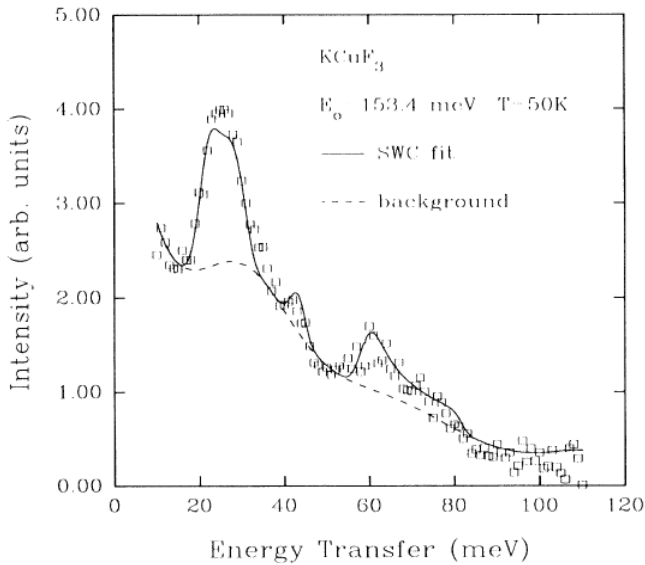

**Figure 4.** Measured scattering in $KCuF_3$ with $E_0$ = 153.4 meV and $k_i$ along c*. The dashed line is an independently determined non-magnetic background, and the solid line is a fit to the background plus magnetic scattering modeled using the Müller ansatz. Figure from Ref. [20].

The initial experiments on $KCuF_3$ were were followed up by additional measurements that applied lessons learned to significantly reduce the non-magnetic background and better optimize the scattering [21,22]. This enabled detailed studies of the temperature dependence of the scattering, that were shown to be in quantitative agreement with field theory calculations based on spinons [23] in the limits of both $q = \pi$ and $q = 2\pi$, clearly establishing the fractional nature of the excitations.

Several additional studies of 1D antiferromagnetic chain systems using MARI followed. The persistence of quantum effects in the model S=3/2 HAFC $CsVCl_3$ [24] showed some evidence both for a renormalization of the spectrum and for increased linewidths. The S=1/2 Ising-like antiferromagnetic chain is another model of interest showing strong quantum effects. The model shows gapped spin-flip excitations that can be thought of as creating pairs of freely moving domain walls ("kinks") producing continuum scattering [25] that is now understood as another manifestation of spinons [26]. At high temperatures thermally excited single walls produce non-lorentzian quasi-elastic scattering showing the dispersive "Villain mode" [27,28,29]. At low temperatures interchain coupling leads to order producing a molecular field leading to a series of sharp bound states known as the "Zeeman ladder" [30, 31]. Applying ToF techniques allowed a more comprehensive investigation of the inelastic scattering in the model material $CsCoCl_3$ [32] than had been previously possible. Finally, a re-examination of the scattering in $CsNiCl_3$ using MARI [33] characterized the continuum scattering in a model S=1 HAFC.

The direct geometry chopper spectrometers were also seen to be excellent instruments for quasi-2D materials, with the quest to understand high-$T_C$ superconductivity providing strong motivation for studying magnetic excitations in several of these systems, including $La_2NiO_4$ [34], $La_2CuO_4$ [35,36] and $La_{1.85}Sr_{0.15}CuO_4$ [37]. In particular the high energy transfers accessible at MARI and HET allowed for a determination of the overall extent of the magnetic fluctuations. Subsequent to these studies a position sensitive detector was installed at HET, facilitating much improved data collection illustrated by additional beautiful measurements on $La_2CuO_4$ [38].

## 3. Dimerized magnetic systems

During the 1990's the inorganic material copper germanate ($CuGeO_3$), with S=1/2 $Cu^{2+}$ ions, received great attention as a prototypical example of an inorganic quasi-1D spin-Peierls system. This is an effect analogous to the electronic Peierls instability, but where the extra exchange energy gained by spontaneously dimerizing compensates for the cost in deformation energy. With continued improvements in the ToF technique at MARI, it was possible to measure the entire spectrum of $CuGeO_3$ with the crystal aligned with $k_i$ perpendicular to the principal chain direction [39], producing a colour contour map showing the full extent of the continuum above and below the transition. In the dimerized state most of the intensity at the lower bound of the spectrum is in a well-defined triplon mode: such modes typically dominate the the spectrum of S=1/2 dimerized systems including alternating chains and ladders. Roger Cowley showed how to apply pseudo-boson methods previously used to understand crystal-field levels to calculate the spectrum of excitations for dimerized magnets [40]. Every local dimer has a singlet ground state, with a creation operator defined for each of the triplet of excited magnetic states. Coupling to the neighbours results in a gapped triplet dispersion relation. This proved a very insightful basis and was used to understand the excitation spectrum in CuGeO$_3$ [40], as well as other materials including CuWO$_4$ [41,42].

## 4. Quantum criticality mapped with MAPS

The ability to carry out ToF neutron spectroscopy on single crystals was revolutionized in the year 2000 when the MAPS spectrometer [43] began operation. MAPS was constructed with an array of position sensitive detectors enabling rapid measurements with much greater coverage in energy and momentum space. This was put to good advantage with numerous measurements of excitations in cuprate materials, including the spinon continuum spectrum of SrCuO$_2$ [44].

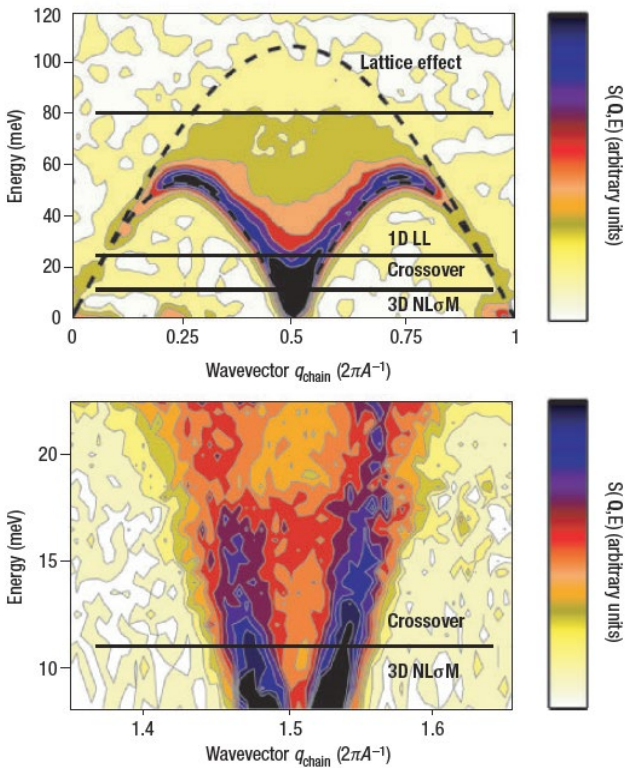

**Figure 5.** (top) Inelastic neutron scattering in KCuF$_3$ at T = 6K measured on MAPS. The various physical regimes are labeled. (bottom) Close up of the low energy region. The red feature near 17 meV is the longitudinal mode, i.e. Higgs mode. (Figure from Ref. [45].

The improved capabilities allowed for a much more detailed investigation of the physics of coupled S=1/2 HAFC as manifest in KCuF$_3$ [45]. The isolated S=1/2 HAFC is can be described as a critical 1D Luttinger Liquid (LL) in the long wavelength limit. The spectrum of fluctuations of the 1D LL is the triply degenerate spinon continuum, and shows E/T scaling (E = energy) as the temperature is varied. For systems that are quasi-1D, even infinitesimal interchain coupling of the chains leads to long range order, and the point where interchain coupling vanishes is a quantum critical point. The spectrum of the coupled chains (Fig.5) shows several distinct regions [46,47]: at very low energies near the antiferromagnetic point one observes sharp transverse spin waves representative of the magnetic order. This region is described in field theory by the 3D non-linear sigma model (NLσM). At high energies characteristic of the 1D interaction strength the response is dominated by the triply degenerate quantum fluctuations of the 1D LL. In between there is a cross-over region exhibiting a longitudinal mode, effectively corresponding to zero-point fluctuations of the ordered moment size. Such a mode had been observed previously in KCuF$_3$ using triple axis methods [48] with its longitudinal character definitively verified via polarization analysis [49]. We note in passing that today such longitudinal modes are referred to as Higgs modes, and remain a subject of active investigation [e.g. 50, 51].

The MAPS data yielded a complete picture of the excitations as a function of momentum, energy and temperature, allowing for a construction of a cross-over diagram showing where one expects to see quantum critical fluctuations [45]. The excellent data quality ultimately enabled quantitative comparisons of the scattering with theoretical calculations going beyond the Müller ansatz using the most modern methods including Bethe ansatz and density matrix renormalization group [52].

## 5. Spin ladders

Spin ladders consisting of coupled S=1/2 chains have been of longstanding interest in part because of predictions that when doped they have a tendency towards superconductivity [53]. Inelastic neutron scattering was used to study a particularly interesting cuprate ladder material La$_4$Sr$_{10}$Cu$_{24}$O$_{41}$ [54]. Strong cyclic exchange was found coming from higher-order hopping processes in the Hubbard Hamiltonian. This takes the ladders closer to quantum criticality. The MAPS spectrometer made these measurements possible due to its high resolution, energy scale, and pixelation. Equally important were the developments in theory of scattering cross sections where the full excitation spectrum including multi-magnon modes were by that time starting to be calculated [55]. Figure 6 shows a close comparison of theory and experiment.

Even more remarkable was the unexpected discovery of a new quantum critical point in another cuprate ladder material CaCu$_2$O$_3$ [56] Here, a distortion in the rungs changed the balance between cyclic exchange to rung exchange taking the material fortuitously close to the critical value of $J_{cyc}/J_{rung}=1/3$. A gapless excitation spectrum showing scaling behavior confirmed this as the first observation of a Wess Zumino Novikov Witten quantum critical point with conformal charge of C=3/2.

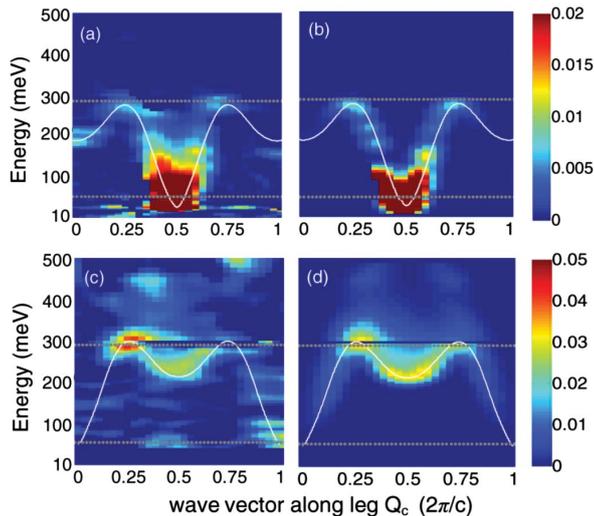

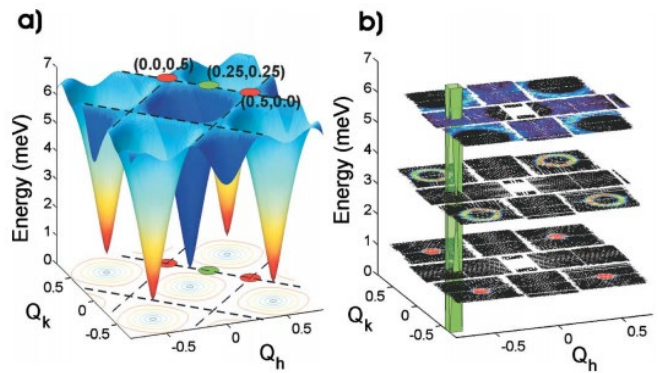

**Figure 6.** Shows the measured scattering from MAPS at ISIS for the ladder material $La_4Sr_{10}Cu_{24}O_{41}$. A parity in the ladder allows odd and even triplon sectors to be separated. (a) Measured one-triplon data with non-magnetic background subtracted. The white curve gives the theoretically calculated one-triplon dispersion curve. (b) Calculated one triplon scattering with corrections for instrumental resolution and magnetic form factor. (c) Measured two-triplon data with nonmagnetic background subtracted. The white curve gives the lower boundary of the two-triplon dispersion curve. (d) Calculated two-triplon scattering. Figure from Ref. 54.

**Figure 7.** One-magnon scattering in $Rb_2MnF_4$. a) The calculated one-magnon dispersion surface as a function of two-dimensional wave vector $Q_h$, $Q_k$ and energy (color shading is intensity in neutron scattering). Dashed lines in the basal plane and at maximum energy $h\omega = 4JS$ mark the antiferromagnetic zone boundaries. The basal plane also shows constant-energy contours (solid lines). b) The constant energy maps of the magnetic scattering at $h\omega = 1, 3.5,$ and 6 meV obtained by taking slices from the 3D $Q_h$, $Q_k$, E neutron data. (Adapted from Ref. 57).

## 6. $Rb_2MnF_4$: observation of two-magnon scattering and quantum classical crossover

The high pixilation of MAPS opened unprecedented possibilities for quantitative analysis, but it was clear that realizing this potential would need more sophisticated modelling and analysis. Roger proposed to measure a material that was well studied and use it to push the technical envelope. It was decided to revisit $Rb_2MnF_4$, a model 2D Heisenberg antiferromagnet. Measurements were undertaken on an excellent single crystal with the incident beam perpendicular to the magnetic layers [57]. Given the very small interplane dispersion, the measurements gave a complete mapping of the excitations in 2D momentum-energy space (see Figure 7). This data set was a prelude to the full 4D mapping techniques that were developed more recently and demonstrated a level of detail and comprehensive coverage of dynamics and thermal effects that was unprecedented at the time.

Developing sophisticated modelling was essential for this problem. Detailed comparison of the low temperature data sets to spin-wave theory was undertaken. These involved full resolution simulations and the treatment of line-shapes. Correcting for pixilation was found to be challenging and a comprehensive treatment of resolution effects in regions of steep dispersion proved essential for the extraction of linewidths [58]. Careful modeling resulted in a textbook demonstration of spin-wave theory [57]. While a highly accurate spin-wave dispersion was extracted from the data, the main achievement was quantitative confirmation of the two-magnon cross section. As $Mn^{2+}$ in $Rb_2MnF_4$ carries a spin-5/2 moment the zero-point fluctuations are relatively small whilst the two-magnon continuum is spread over a large area in wavevector and energy resulting in a signal two orders of magnitude less intense than the one magnon branches [57]. Moreover the scattering intensity depends on the experimental configuration, including the crystal orientation. Figure 8 shows an example of the two magnon signal and comparison to calculations.

The rapid measurements possible at MAPS allowed the temperature dependence of the scattering to be investigated in detail. The data was compared to spin simulations carried out using Landau-Lifshitz dynamics [58,59]. This yielded the insight that a crossover occurs at a temperature T* equal to the Curie-Weiss temperature $\Theta_{CW}$ normalized by the single-site spin value S, i.e. $T^* = \Theta_{CW}/S$ [59]. For $T > T^*$ the data can be quantitatively explained by classical dynamics, with quantum effects becoming important for $T < T^*$. As an aside it is worth mentioning that Landau-Lifshitz dynamics and quantum classical crossover ideas are enjoying renewed interest in connection with quantum spin liquid research. There semiclassical methods are proving to be effective at calculating scattering cross sections at finite temperature [60,61,62].

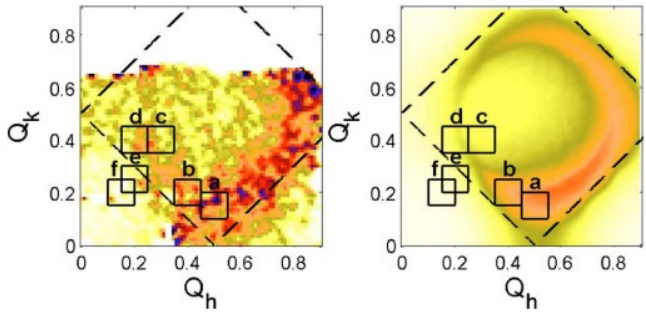

**Figure 8.** An example of a comparison between data and predicted two-magnon scattering for $Rb_2MnF_4$ at an energy much higher than the one-magnon zone boundary energy. The simulated two-magnon scattering intensity (designated by color) includes the polarization factor and magnetic form factor. The dashed square box is the antiferromagnetic zone boundary. Boxes labeled (a)–(f) show the locations of energy scans explained in detail in Ref. 57. (Adapted from Ref. 57).

Another significant result of the experiments on $Rb_2MnF_4$ was the observation that the wave-vector dependence of the spin-wave damping was not in agreement with existing theories [63,64]. Subsequent high-resolution triple-axis resonance spin echo measurements confirmed this and showed that the disagreement could be fixed with improved calculations, resolving questions that had persisted over several decades [65].

## 7. Cold pulsed neutron spectroscopy and the quasi-2D frustrated quantum magnet $Cs_2CuCl_4$

Some of the most interesting problems in low D quantum magnetism involved small energy scales that were not well-matched to the available direct geometry chopper spectrometers. This led researchers to utilize cold-neutron indirect geometry instruments originally designed for molecular spectroscopy. The IRIS and OSIRIS instruments at ISIS offered resolutions better than 30 μeV and adequate wavevector resolution. Extensive work with the sample environment was needed and software to plan, visualize, and analyze the experiments, but in the end the data quality was quite remarkable [66] and represented a new level of measurement of continua in quantum systems on milli-eV energy scales.

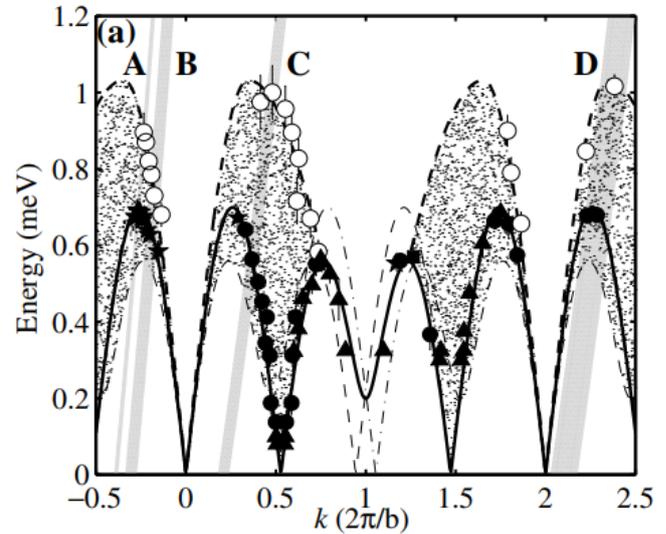

**Figure 9.** boundaries of dispersive continua of scattering measured in $Cs_2CuCl_4$ at $T=0.1$ K on the IRIS spectrometer. The panels are from different regions of the 2D plane. Filled symbols are the main peak in the line shape and the solid line is a fit to a dispersion relation. The dotted area indicates the extent of the magnetic scattering, large open circles mark the experimentally estimated continuum upper boundary and the upper thick dashed line is a guide to the eye. Open squares indicate the lower boundary of a second continuum. (From Ref. 70).

One of the motivating problems was the desire to more directly observe the band nature of spinons by using magnetic fields to split the spin fluctuations into incommensurate continua. The fields required to induce these effects in $KCuF_3$ were hundreds of Tesla which was impossible to reach. In fact, very few low-energy materials were available as large single crystals at the time, especially without hydrogen - the cause of a considerable background from incoherent scattering. In addition, the desire to reach saturation, where quantum fluctuations would be suppressed and unrenormalized exchange couplings measurable provided stringent constraints. Quantum renormalization effects were known to be large but hadn't been directly measured before. The material $Cs_2CuCl_4$ [67] was identified from magnetic susceptibility as being suitable and Zbigniew Tylczynski in Poland supplied excellent single crystals.

The low exchange interaction required working at dilution temperatures and combining this with magnetic fields was technically very difficult, but the measurements were pursued anyway. After two demanding experiments at the ILL, field shifting of the continua was observed [68] in line with calculations [14,69].

The measured continua of scattering showed incommensuration even without field with strong differences from a chain [70] (see Fig. 9). This was identified as being due to the frustrated triangular interchain coupling. In fact, this coupling was far from insignificant as we determined by measuring in very high fields of 15 Tesla (at V2 Hahn Meitner Institute) with $Cs_2CuCl_4$ in a fully magnetically saturated state [71]. The excited states in the saturated phase are simple magnons and the couplings straightforwardly determined. The interchain coupling was near 1/3$^{rd}$ of the chain coupling and Cs2CuCl4 then was an anisotropic 2D rather than a 1D material [71]. As a by-product of this measurement, a detailed experiment of Bose-Einstein condensation of magnon modes was also possible showing the link between dynamical condensation and ground state formation [71]. The existence of 2D dispersive continua that were observed in Cs2CuCl4 were characteristic of a spin liquid with resonating valence bonds singlets [70]. These highly quantitative studies attracted great interest and Balents and coworkers in the end formulated a comprehensive theory of the magnetism in terms of confined spinons [72].

## 8. $CoNb_2O_6$: Ising Model in a Transverse Field

The application of quantum field theories to neutron data provides a powerful approach to understanding strongly fluctuating quantum states. The use of milliKelvin temperatures in conjunction with magnetic fields allowed quantum effects to be studied in different phases and at quantum phase transitions. The first way to apply fields was as a chemical potential via the Zeeman energy to drive quasiparticles into ground states as had been successfully employed in $Cs_2CuCl_4$. A second was to use the magnetic field in non-commuting spin direction and so to drive quantum phase transitions by tuning the quantum tunnelling. These methods were applied to two materials, $Cs_2CoCl_4$ [73] and $CoNb_2O_6$ [74]. In the case of $CoNb_2O_6$, which was a near ideal example of a ferromagnetic Ising chain, a field of 6T proved to be enough to quantum disorder the material and drive the quantum phase transition from an ordered to quantum disordered state. The measurements were made on the OSIRIS instrument, which unlike IRIS, was able to take magnetic fields without disrupting its detectors. The experiments led to a textbook study of a quantum phase transition and showed quantum critical exponents in the region of the critical point. A remarkable surprise was the detection, due to the very high resolution of the instrument, of a series of multiparticle bound states forming in the continuum in the ordered phase due to confinement effects on domain wall pairs by interchain coupling. This is in some ways similar to the Zeeman ladder seen in Ising-like antiferromagnets [30,31, 32] but with a confining potential controlled by the applied field. In low field modes can be well described by a Schroedinger model of domain walls in a potential well. As the quantum critical point was approached these modes show remarkable behaviour where they form into bound modes of each other. The ratios between their energies and intensities are predicted to be given by the E8 exceptional Lie group, a hidden property of the quantum critical point and indeed the measurements show the modes conforming to this remarkable symmetry.

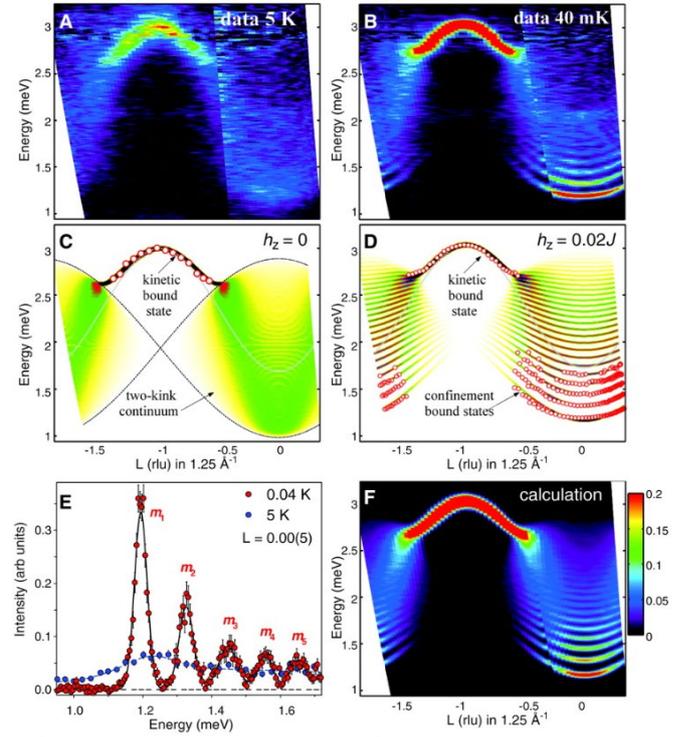

**Figure 10**. Zero-field spin excitations observed by neutron scattering in $CoNb_2O_6$. (A) At 5 K in the 1D phase above $T_N$ a broad continuum occurs due to scattering by pairs of kinks that propagate independently on the thermally-decoupled chains. (B) The continuum splits into a "Zeeman ladder" of resonances deep in the 3D ordered phase (0.04 K) due to confinement of kinks by the weak interchain couplings. (C) and (D) Schematics of the model calculations without and with order for comparison with data in (A) and (B). (E) shows the resonance lines in the ordered phase and (F) is a full calculation from solution of the two-kink Schroedinger equation of the scattering cross section quantitatively describing the data. Figure from Ref. 74.

# 9. Some modern applications of pulsed neutron spectroscopy

The success of pulsed neutron spectroscopy spurred the development of more advanced instruments and sources. Today researchers can use, among others, MAPS, MERLIN and LET at ISIS, ARCS, SEQUOIA, HYSPEC and CNCS at the Oak Ridge National Laboratory Spallation Neutron Source (SNS), or the 4SEASONS spectrometer at the Japan Proton Accelerator Research Complex (JPARC). Large position-sensitive detector arrays are now standard, and there have been significant technical advances in both computation and data storage. Early generations of pulsed neutron inelastic spectrometers generally collected and stored data as histograms, adding up the counts from some (typically large) number of pulses. Typically for each pulse in an inelastic scattering measurement most time channels record zero counts. Therefore it can be more efficient to store data in event mode, where each counted neutron is tagged with all relevant information connected with the sample condition as well as time bin, scattering angles, etc. This information can be used to reconstruct the scattering in momentum-energy space, or as a function of real-time, temperature, or any other recorded variable. Inelastic experiments on single-crystals now typically collect data for numerous crystal orientations, fully covering a defined volume in 4D momentum energy space.

The interest in many different low D magnetic systems continues unabated, and in recent years pulsed neutron spectroscopy, has been applied to measurements of spinon spectra in quasi-1D rare-earth based systems [75,76], dimer magnets showing magnetic field driven Bose-Einstein condensation of triplons [77], and, as mentioned earlier, Higgs modes [50]. However, in recent years much of the interest has focused on excitations in 2D materials related to quantum spin liquids [representative recent reviews include 78, 79].

Quantum spin liquids (QSLs) are characterized by the lack of magnetic order down to zero temperature, strong entanglement, and the presence of fractional magnetic excitations. The QSL systems considered first were $S=1/2$ systems with isotropic Heisenberg interactions on frustrated 2D lattices [80]. Along these lines neutron scattering evidence for continuum scattering characteristic of fractional excitations in a QSL was seen in the material Herbertsmithite [81], a realization of the $S=1/2$ Heisenberg antiferromagnet on a Kagomé lattice. Much recent interest though, has turned to systems with anisotropic interactions, perhaps exemplified by the Kitaev model on a honeycomb lattice (KHL) [82].

The KHL model consists of $S=1/2$ spins interacting with their nearest neighbours via Ising exchange, with the stipulation that the Ising anisotropy direction is parallel to the bond connecting the spins. Thus each spin experiences three mutually incompatible interactions, leading to a fully frustrated system for either ferromagnetic or antiferromagnetic interactions. The KHL is exactly solvable, and the ground state is rigorously a QSL, with fractional excitations consisting of static gauge fluxes (also known as visons) and mobile Majorana fermions [82]. The expected dynamical susceptibility of the model can also be calculated exactly [83]. The model attracted great attention since in principle the Majoranas could be the basis for a topologically protected qubit. It was subsequently shown how the terms in the KHL model could be realized in materials [84], and these developments have been extensively reviewed [85, 86, 87].

The initial search for material realizations of the Kitaev QSL focused on iridate materials but it was later realized that the magnetic insulator $\alpha$-RuCl$_3$ was also a reasonable candidate [88,89]. Although the material orders at low temperature (7K in a sample with minimal stacking faults), inelastic neutron scattering on powder samples gave strong evidence that it was proximate to a QSL, possibly that related to the Kitaev model [89]. Measurements on single crystals using direct geometry spectrometers [90, 91] showed the scattering to be very unusual: spin wave excitations at low temperature coexisted with a broad continuum, giving constant energy scans a shape in reciprocal space resembling the star of David at low temperatures (see Fig.11). The continuum scattering persists to high temperatures and interestingly has roughly the extent in momentum-energy space expected for the response of a Kitaev QSL.

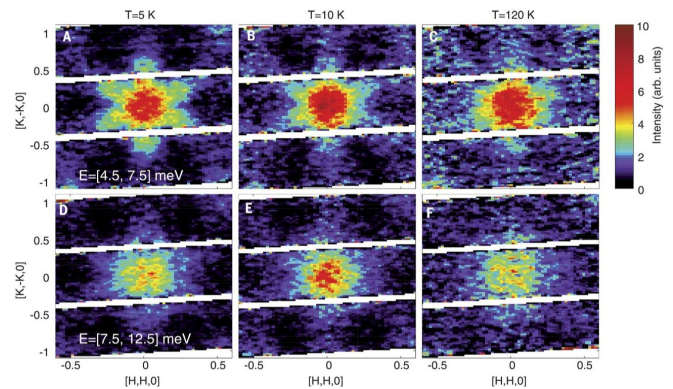

**Figure 11.** Scattering continuum in $\alpha$-RuCl$_3$, measured using the SNS SEQUOIA spectrometer. Neutron scattering measurements using fixed incident energy $E_i$ = 40 meV, projected on the reciprocal honeycomb plane defined by the perpendicular directions ($H$, $H$, 0) and ($K$, –$K$, 0), integrated over the interval $L$ = [–2.5, 2.5]. Intensities are denoted by color, as indicated in the scale at right. Measurements integrated over the energy range [4.5, 7.5] meV are shown on the top row at temperatures (**A**) 5 K, (**B**) 10 K, and (**C**) 120 K. The corresponding measurements integrated over the interval [7.5, 12.5] meV are shown on the bottom in (**D**), (**E**), and (**F**). The white regions lack detector coverage. Figure from Ref. 90.

Applying a sufficient magnetic field in the honeycomb plane of α-RuCl$_3$ suppresses the magnetic order, possibly leading to a QSL state [92]. ToF inelastic neutron scattering in an applied field was carried out using the HYSPEC instrument, showing that at the critical field the spin wave scattering vanishes and the continuum, presumed to be the signature of fractional excitations, is enhanced [93]. The possibility that these fractional excitations are indeed Majorana fermions was bolstered by the observation of a quantized thermal Hall effect over an interval of applied field with an onset coinciding with the suppression of magnetic order [94]. As of this writing this result has not been independently verified, but recent inelastic neutron scattering and magnetocaloric effect measurements [95] point to a temperature-field phase diagram compatible with that inferred from the thermal Hall measurements.

Intrinsic disorder can play a very important role in affecting real or apparent QSL behavior, as was shown in a recent study of the triangular lattice system YbMgGaO$_4$ [96] using CNCS. Many other potential QSL systems are now under active investigation and one expects that ToF inelastic neutron scattering will be an indispensable tool for sorting out the physics of candidate materials.

## 10. The future

The study of quantum phases and their excitations has grown dramatically in recent years. Much of the remarkable physics of Berry's phases, topological states and excitations, and quantum criticality that was explored using neutrons on low dimensional magnets has found much wider application as acknowledge by the award of the Nobel prize to Thouless, Haldane and Kosterlitz. An incredible range of quantum phenomena are now being studied with inelastic TOF from Majorana fermions to magnetic monopoles and emergent photons in spin liquids to topological magnons and Weyl semimetals.

TOF neutron scattering has become an essential technique in experimental science. An array of powerful pulsed neutron sources are available worldwide with inelastic scattering from single crystals as a central part of their mission. At the same time, reactor-based instrumentation has increasingly used large detector banks using TOF techniques and multi crystal analysers.

Plans for a new generation of sources and instrumentation are in hand. The European Spallation Source in Lund, Sweden and the Second Target Station at the Spallation Neutron Source, Oak Ridge National Laboratory, will deliver a new level of neutron scattering capabilities particularly in the cold neutron regime. These sources and their planned instruments promise orders of magnitude increase in measurement performance and will open up the possibility of working on excitations in heterostructures, pump-probe and out-of-equilibrium experiments, mapping of Q,E space with the full polarization tensor, as well as measurements on much smaller samples, breaking through a critical barrier in the need for large single crystals for experiments.

Developments in computation have played a central role in opening up the study of excitations with neutrons. High performance computing is opening the possibility of a new level of simulations of the dynamics in quantum magnets and machine learning is just starting to be applied which is showing very promising results [97].

## 11. Summary remarks

Overall the contribution to the development of time-of-flight techniques by Roger Cowley, his and our collaborators, and group in Oxford was driven by research into key questions in quantum magnetism that have much impact on contemporary research. The proposals from theory for exotic quantum states in low dimensional magnets, particularly from quantum field theories that pointed to a potentially rich vein of new physics have more than fulfilled their promise. All along the challenge was to develop experimental capabilities to observe and manipulate these phases and so bring confirmation of their existence as well as the chance to make discoveries. This has been achieved and many new effects have been found.

Although in principle ideal for probing quantum magnets, there were many experimental challenges in measurement, such as working at milliKelvin temperatures and in magnetic fields to realize the potential of neutrons as a probe. Often the 1 quasiparticles arising in quantum magnets gave rise to unusual scattering signals that conventional neutron techniques were not designed to address. ToF techniques offered the possibility of mapping out weak continua of excitations and giving the energy resolution to measure singularities in scattering cross sections indicative of strong correlations.

The challenges in developing the experimental technique were formidable and required close interaction with theory. This was because thoroughly understanding the relationship of measured correlation functions to underlying states was central to extracting the physics. In this respect the developments in Oxford benefitted greatly from a wealth of theoretical talent with Alexei Tsvelik, Fabian Essler, Jean-Sebastien Caux, John Chalker and Roderich Moessner among others. Further, nearby access to ISIS meant that one was able to develop the software and instrumentation needed to make the measurements and analyze them. In fact, the fundamental step of using sophisticated algorithms to handle data and simulate the scattering was crucial. Roger correctly realized that a change from algebraic to algorithmic thinking was important and supported the extensive computational needs.

Most important of all, Roger attracted a cohort of talented students and postdocs who worked on opening up this area of science, many of whom have gone on to distinguished careers. Examples include Radu Coldea, Jon Goff, Michel Kenzelmann, Bella Lake, and Des McMorrow.

**Acknowledgements**

A portion of this research used resources at the High Flux Isotope Reactor and Spallation Neutron Source, DOE Office of Science User Facilities operated by the Oak Ridge National Laboraory.

**References**

[1] e.g. Mermin N. D. and Wagner H. 1966 *Phys. Rev. Lett.* **17** 1133
[2] de Jongh, L. J. and Miedema A.R. 1974 *Adv. Phys.* **23** 1
[3] Steiner M, Villain J and Windsor C. G. 1976 *Adv. Phys*. **25** 87
[4] Hutchings M. T. *et al.* 1972 *Phys. Rev.* **B5** 1999
[5] Bethe H. A. 1931 *Z. Phys*. **71** 205
[6] Des Cloizeaux J. and Pearson J.J. 1962 *Phys. Rev.* **128** 2131
[7] Endoh Y. *et al.* 1974 *Phys. Rev. Lett.* **32** 170
[8] Heilmann I.U. *et al.* 1978 *Phys. Rev.* **B18** 3530
[9] Yamada T. 1969 *Prog. Theor. Phys.* **41** 880
[10] Bonner J. C. Sutherland B. and Richards P. M. 1975 *AIP Conf.* **24** 355
[11] Faddeev L. D. and Takhtajan L. A. 1981 *Phys. Lett.* **85A** 375
[12] Haldane F. D. M. 1983 *Phys. Lett.* **93A** 464. An earlier 1981 version was unpublished but available at arXiv:1612.00076.
[13] Buyers W. J. L. *et al.* 1986 *Phys. Rev. Lett.* **56** 371
[14] Müller G. *et al.* 1981 *Phys. Rev.* **B24** 1429
[15] Perring T. G. 1991 Ph. D. Thesis *High Energy Magnetic Excitations in Hexagonal Cobalt* Cambridge University.
[16] Taylor A.D. *et al.* 1991 *Proceedings of ICANS XI, KEK Report 90-25* 705
[17] Welz D. *et al.* 1992 *Phys. Rev.* **B45** 12319
[18] Hutchings M.T. Ikeda H. and Milne J.M. 1979 *J. Phys. C* **12** L739
[19] Satija S.K. *et al.* 1980 *Phys. Rev.* **B21** 2001
[19] Taylor A.D. *et al.* 1991 *Proceedings of ICANS XI, KEK Report 90-25* 705
[20] Nagler S.E. *et al.* 1991 *Phys. Rev.* **B44** 12361
[21] Tennant D. A. *et al.* 1993 *Phys. Rev. Lett.* **70** 4003
[22] Tennant D. A. *et al.* 1995 *Phys. Rev.* **B52** 13368
[23] Schulz H. J. 1986 *Phys. Rev.* **B34** 6372
[24] Itoh S. *et al.* 1995 *Phys. Rev. Lett.* **74** 2375
[25] Ishimura N. and Shiba H. 1980 *Prog. Theor. Phys.* **63** 743
[26] Bougourzi A. H., Karbach M., and Müller G 1998 *Phys. Rev.* **B57** 11429
[27] Villain J. 1975 *Physica* **79B** 1
[28] Yoshizawa H. *et al.* 1981 *Phys. Rev.* **B23** 2298
[29] Nagler S. E. *et al.* 1982 *Phys. Rev. Lett.* **49** 590
[30] Shiba H. 1980 *Prog. Theor. Phys.* **64** 466
[31] Nagler S. E. *et al.* 1983 *Phys. Rev.* **B27** 1784
[32] Goff J. Tennant D. A. and Nagler S. E. 1995 *Phys. Rev.* **B52** 15992
[33] Kenzelmann M. *et al.* 2001 *Phys. Rev. Lett.* **87** 017201
[34] Yamada K. *et al.* 1991 *J. Phys. Soc. Japan* **60** 1197
[35] Hayden S.M. *et al.*, 1991 *Phys. Rev. Lett.* **67** 3622
[36] Itoh S. *et al.* 1994 *J. Phys. Soc. Japan* **63** 4542
[37] Yamada K. *et al.* 1995 *J. Phys. Soc. Japan* **64** 2742
[38] Coldea R. *et al.* 2001 *Phys. Rev. Lett.* **86** 5377
[39] Arai M. *et al.* 1996 *Phys. Rev. Lett*. **77** 3649
[40] Cowley R.A. Lake B. and Tennant D.A. 1996 *J. Phys. Cond. Matt*. **8** L179
[41] Lake B. *et al.* 1996 *J. Phys. Cond. Matt.* **8** 8613
[42] Lake B. Cowley R.A. and Tennant D.A. 1997 *J. Phys. Cond. Matt.* **9** 10951
[43] Perring T.G. et al. 1994 *Proceedings of ICANS XII* I-60
[44] Zaliznyak I. A. *et al.* 2004 *Phys. Rev. Lett.* **93** 087202
[45] Lake B. *et al.* 2005 *Nature Materials* **4** 329
[46] Schulz H.J. 1996 *Phys. Rev. Lett.* **77** 2790
[47] Essler F.H.L. Tsvelik A.M. and Delfino G. 1997 *Phys. Rev.* **B56** 11001
[48] Lake B. Tennant D.A. and Nagler S. E. 2000 *Phys. Rev. Lett.* **85** 832
[49] Lake B. Tennant D.A. and Nagler S. E. 2005 *Phys. Rev.* **B71** 134412
[50] Jain A. *et al.* 2017 *Nat. Phys.* **13** 633
[51] Hong T. *et al.* 2017 *Nat. Phys.* **13** 638
[52] Lake B. *et al.* 2013 *Phys. Rev. Lett.* **111** 137205
[53] Dagotto E. Riera J. and Scalapino D. 1992 *Phys. Rev.* **B45** 5744
[54] Notbohm S. *et al.* 2007 *Phys. Rev. Lett.* **98** 027403
[55] Schmidt K.P. *et al.* 2003 *Phys. Rev. Lett.* **90** 227204
[56] Lake B. *et al.* 2010 Nat. Phys. **6** 50
[57] Huberman T. *et al.* 2005 *Phys. Rev.* **B72**, 014413
[58] Huberman T. 2004 Ph. D. thesis *Neutrons scattering studies and classical simulations on low-dimensional spin systems* University of Oxford
[59] Huberman T. *et al.* 2008 *J. Stat. Mech.: Theory and Exper.* P05017
[60] Samarakoon A. *et al.* 2017 *Phys. Rev.* **B96** 134408
[61] Samarakoon A. *et al.* 2018 *Phys. Rev.* **B98** 045121
[62] Zhang S. *et al.* 2019 *Phys. Rev. Lett.* **122** 167203
[63] Harris A. *et al.* 1971 *Phys. Rev.* **B3** 961
[64] Tyc S. and Halperin B. 1990 *Phys. Rev.* **B42** 2096
[65] Bayrakci S. *et al.* 2013 *Phys. Rev. Lett.* **111** 017204
[66] Coldea R. Tennant D.A. and Tylczynski Z. 2003 *Phys. Rev.* **B68** 134424
[67] Coldea R. *et al.* 1996 *J. Phys. Cond. Matt.* **8** 7473
[68] Coldea R. *et al.* 1997 *Phys. Rev. Lett.* **79** 151
[69] Talstra J.C. and Haldane F.D.M. 1994 *Phys. Rev.* **B50** 6889
[70] Coldea R. *et al.* 2001 *Phys. Rev. Lett.* **86** 1335
[71] Coldea R. *et al.* 2002 Phys. Rev. Lett. **88** 7203
[72] Kohno M. Starykh O.A. and Balents L. 2007 *Nature Physics* **3** 790
[73] Kenzelmann M. *et al.* 2002 *Phys. Rev.* **B65** 4432
[74] Coldea R. *et al.* 2010 *Science* **327** 177
[75] Wu L.S. *et al.* 2016 *Science* **352** 1206
[76] Wu L.S. *et al.* 2019 *Nat. Comm.* **10** 698
[77] Hester G. *et al.* 2019 *Phys. Rev. Lett.* **123** 027201
[78] Savary L. and Balents L. 2017 *Rep. Prog. Phys.* **80** 016502.
[79] Knolle J. and Moessner R. 2019 *Ann. Rev. Cond. Matt. Phys.* **10** 451
[80] Anderson P.W. 1973 *Mater. Res. Bull.* **8** 153
[81] Han T.-H. *et al.* 2012 *Nature* **492** 406
[82] Kitaev A. 2006 *Ann. Phys.* **321** 2
[83] Knolle J. *et al.*, 2014 *Phys. Rev. Lett.* **112** 187201

[84] Jackeli G. and Khaliullin G. 2009 *Phys. Rev. Lett. 102* 107205
[85] Winter S.M. *et al*. 2017 *J. Phys. Cond. Matt*. **29** 493002
[86] Hermanns M. Kimchi I. and Knolle J. 2018 *Ann. Rev. Cond. Matt. Phys*. **9** 17
[87] H. Takagi et al. 2019 *Nature Reviews Physics* 1 264
[88] Plumb K.W. *et al.* 2014 *Phys. Rev.* **B90** 041112(R)
[89] Banerjee A. *et al*. 2016 Nat. Mater. **15** 733
[90] Banerjee A. *et al*. 2017 *Science* **356** 1055
[91] Do S.-H. *et al*. 2017 *Nat. Phys.* **13** 1079
[92] Baek S.-H. *et al*. 2017 *Phys. Rev. Lett.* **119** 037201
[93] Banerjee A. *et al*. 2018 *npj Quantum Materials* **3** 8
[94] Kasahara Y. *et al.* 2018 *Nature* **559** 227
[95] Balz C. *et al.* 2019 *Phys. Rev.* **B100** 060405
[96] Paddison J.A.M. *et al.* 2017 *Nat. Phys.* **13** 117.
[97] Samarakoon A. *et al.* 2019 *Nat. Commun.* to be published